\documentclass[11pt,cite]{article}
\usepackage{latexsym}

\usepackage[latin1]{inputenc}

\usepackage{graphicx}
\usepackage{color,pst-plot}

\usepackage{amsmath}
\usepackage {amsfonts,amssymb,amstext}

\newcommand{\lbl}[1]{\label{eq:#1}}
\newcommand{ \rf}[1]{(\ref{eq:#1})}

\newcommand{\be}{\begin{equation}}
\newcommand{\ee}{\end{equation}}
\newcommand{\bea}{\begin{eqnarray}}
\newcommand{\eea}{\end{eqnarray}}
\newcommand{\setl}{\setlength\arraycolsep{2pt}}

\newcommand{\noi}{\noindent}
\newcommand{\nn}{\nonumber}
\newcommand{\ra}{\rightarrow}

\newcommand{\cA}{{\cal A}}

\newcommand{\cO}{{\cal O}}

\newcommand{\Imm}{\mbox{\rm Im }}
\newcommand{\Ree}{\mbox{\rm Re }}

\newcommand{\Nc}{\mbox{${\rm N_c}$}}

\input epsf

\def\theequation{\arabic{section}.\arabic{equation}}

\begin{document}

\begin{titlepage}

\begin{flushright}
\today  \\
LAPTH-024/13
\end{flushright}

\vspace*{0.2cm}
\begin{center}

{\Large{\bf The Froissart--Martin Bound for $\pi\pi$ Scattering in QCD}}
\\[2 cm]

{
{\bf David Greynat}~$^{a,b}$ and {\bf Eduardo de Rafael}~$^c$}\\[1cm]

$^a$ {\it  LAPTh.\\
 Université  de Savoie, CNRS\\
 B.P.110, Annecy-le-Vieux F-74941, France} \\[0.5cm]

$^b$ {\it Dipartimento di Scienze Fisiche, Universita  di Napoli "Federico II"\\
 Via Cintia, 80126 Napoli, Italia} \\[0.5cm]

$^c$  {\it Aix-Marseille Universit\'e, CNRS, CPT, UMR 7332, 13288 Marseille, France\\
   Université de Toulon, CNRS, CPT, UMR 7332, 83957 La Garde, France }
    
\end{center} 

\vspace*{3.0cm}   

\begin{abstract}
The Froissart--Martin bound for total $\pi\pi$ scattering cross sections is reconsidered in the light of QCD properties such as spontaneous chiral symmetry breaking and the counting rules for a large number of colours $\Nc$. 
\end{abstract}

\end{titlepage}

\section{\normalsize Introduction}\lbl{int}
\setcounter{equation}{0}
\def\theequation{\arabic{section}.\arabic{equation}}

\noi 
Since the early work by  Froissart~\cite{Froi61},  Martin~\cite{Mart66} and colleagues~\cite{JM64} have shown in a series of seminal papers~\footnote{See e.g. ref.~\cite{TTWetal11} where earlier references can be found.} that under very general assumptions, total cross sections for $\pi\pi$, $K\bar{K}$, $\pi K$, $\pi N$ and $\pi\Lambda$ scattering cannot grow faster than
\be\lbl{eq:FM}
\sigma^{\mbox{\footnotesize tot}}(s) \underset{{s\ \ra \infty}}{\thicksim}\ \frac{4\pi}{t_0}\log^2 \frac{s}{s_0}\,,
\ee
where $s$ is the total center of mass energy squared,  $t_0$ denotes the lowest mass squared singularity in the $t$--channel, which for the processes mentioned above occurs at $4 m_{\pi}^2$, and the normalization $s_0$ in the $\log^2 s$ is arbitrary indicating where the asymptotic behaviour sets in. 
Although several hadronic models have been shown to saturate the Froissart--Martin (FM) bound~\footnote{See e.g. refs.~\cite{FIIL02, DGN02} and references therein.},
it is quite frustrating that the advent of QCD as the theory of the strong interactions has not added anything new, at least so far, on the FM bound. Some obvious questions which one would like to answer are:

\begin{enumerate}
 
\item What happens in QCD in the chiral limit where pions, the Nambu--Goldstone states of the chiral $SU(2)$ flavour symmetry of QCD, become massless? Does the bound become irrelevant, as the presence of the pion mass in the denominator in Eq.~\rf{eq:FM} seems to indicate?
	
\item What becomes of the FM bound in the Large--$\Nc$ limit of QCD? The Large--$\Nc$ counting rules fix $\sigma^{\mbox{\footnotesize tot}}(s)$ in Eq.~\rf{eq:FM} to be of $\cO\left(1/\Nc\right)$, while the FM--bound appears to be of   $\cO(1)$.
	
\item Independently of the previous questions concerning the chiral limit and the Large--$\Nc$ limit, one would also like to know if  the $\log^2 s$ behaviour of the FM bound is saturated in QCD?
	
\end{enumerate}

The purpose of this paper is to set the path to an investigation of these questions.
Here we shall limit ourselves to the case of  total cross sections for $\pi\pi$ scattering. In the next section we summarize well known properties of the elastic $\pi\pi$ scattering amplitudes which we shall need for our discussion. The framework of our analyses uses a Mellin--Barnes representation for the $\pi\pi$ amplitudes which we present in Sec. III. This will allow us to fix the discussion concerning the first question above. Section IV is dedicated to a discussion of the FM bound within the framework of the QCD Large--$\Nc$ limit. Our conclusions are given in Sec. V.

\vspace*{0.5cm}

\section{\normalsize Elastic Pion--Pion Scattering.}
\setcounter{equation}{0}
\def\theequation{\arabic{section}.\arabic{equation}}

\noi 
Elastic $\pi\pi$ scattering in the the isospin symmetry limit is  described by a single invariant Lorentz amplitude $A(s,t,u)$~\footnote{For a modern review see ref.~{\cite{ACGL01}}.}.

{\setl
\bea
\lefteqn{
\langle \pi^{d}(p_4)\pi^{c}(p_3)\ {\rm out}\vert \pi^{a}(p_1) \pi^{b}(p_2)\ {\rm in}\rangle =}\nn \\
 & & \hspace*{-1cm} {\bf 1}+i(2\pi)^4 \delta^4 (p_3 +p_4 -p_1 -p_2)\left\{ \delta^{ab}\delta^{cd}A(s,t,u) + \delta^{ac}\delta^{bd}A(t,u,s)+\delta^{ad}\delta^{bc}A(u,s,t)\right\}\,,\lbl{eq:amp}
\eea}

\noi
where $a$, $b$, $c$, $d$ denote the 1,2,3 components of the adjoint representation of the pion fields in $SU(2)$ and $s$, $t$ and $u$ the usual Mandelstam variables  constrained by
\be
s+t+u=4m_{\pi}^2\,.
\ee
Because of the optical theorem which relates the absorptive part of an elastic amplitude to a total cross section we 
shall only consider elastic scattering amplitudes with the same in and out quantum numbers, i.e.:

{\setl
\bea
\cA_{\pi^\pm \pi^0 \ra \pi^\pm \pi^0}(s,t) & = & A(t,u,s)\,,\nn\\
\cA_{\pi^0 \pi^0 \ra \pi^0 \pi^0}(s,t) & = & A(s,t,u)+ A(t,u,s)+A(u,s,t)\,,\nn\\ 
\cA_{\pi^+ \pi^- \ra \pi^+ \pi^- }(s,t) & = & A(s,t,u)+ A(t,u,s)\,,\nn\\
\cA_{\pi^{\pm} \pi^{\pm} \ra \pi^{\pm} \pi^{\pm}}(s,t) & = & A(t,u,s)+A(u,s,t)\,.
\eea}

\noi

It is convenient to work with the three $s$--channel isospin components ${\bf T}=(T^0, T^1, T^2)$ of the amplitudes in Eq.~\rf{eq:amp}  given by:

{\setl
\bea
T^0 (s,t) & = & 3 A(s,t,u)+A(t,u,s)+A(u,s,t)\,,\nn\\
T^1 (s,t) & = & A(t,u,s)-A(u,s,t)\,,\nn\\
T^2 (s,t) & = & A(t,u,s)+A(u,s,t)\,.
\eea}

\noi 
These amplitudes obey fixed-$t$ dispersion relations, valid in the interval $-28m_{\pi}^2 <t< 4 m_{\pi}^2$. They are the so called Roy equations~{\cite{Roy71}} which we shall consider at $t=0$ and, because of our first question in the Introduction concerning the chiral limit of the FM bound, at $m_{\pi}\ra 0$.
The Roy equations simplify then as follows:

{\setl
\bea\lbl{eq:roy}
\lefteqn{\Ree
\left(\begin{array}{c}
{T}^0 (s,0)\\
{T}^1 (s,0)\\
{T}^2 (s,0)
\end{array}\right)=\frac{s}{f_{\pi}^2} \left(\begin{array}{c} ~~2 \\ ~~1 \\ -1
\end{array}~~\right)}\nn \\
& & \hspace*{-1cm} +s^2 \int_0^{\infty}ds' \frac{1}{s'^2}\left[\frac{1}{s'-s}\left(\begin{array}{ccc}1 & 0 & 0 \\
0 & 1 & 0 \\
0 & 0 & 1\end{array} \right) +\frac{1}{s'+s}\left(\begin{array}{ccc}1/3 & -1 & 5/3\\
-1/3 & 1/2 & 5/6 \\
1/3 & 1/2 & 1/6 \end{array} \right)\right]\frac{1}{\pi}\Imm \left(\begin{array}{c}
{T}^0 (s',0)\\
{T}^1 (s',0)\\
{T}^2 (s',0)
\end{array}\right)\,.
\eea}

\noi
The term in the r.h.s. in the first line of this equation reflects the two subtractions which have been made, as required by the Froissart bound~\footnote{Notice, however,  that the presence of the two powers of $\log s$ in the asymptotic behaviour of the absorptive amplitudes does not restrict any further the two subtractions which are already required for a cross section going as a constant at $s\ra\infty$ .}. In QCD, the explicit values of these subtractions are fixed by lowest order $\chi$PT~\cite{Wei79}. We recall that in chiral $SU(2)$  the amplitude $A(s,t,u)$ in the limit we are considering is given by the $\chi$PT expansion (see ref.~\cite{GL84} and earlier references therein):

{\setl
\bea\lbl{eq:GLAP}\lefteqn{
A(s,t,u) 
\underset{{s,\ t,\ u\ \ra\  0}}{\thicksim}\  \frac{s}{f_{\pi}^2}+\frac{1}{f_{\pi}^4}\left[2 s^2\ l_1 ^{\rm r}+\left[s^2 +(t-u)^2\right]\frac{1}{2}\ l_2 ^{\rm r} \right]}\nn \\
 &  & +  \frac{1}{96\pi^2 f_{\pi}^4}\left[3s^2 \left(\log\frac{\mu^2}{-s}+\frac{5}{6} \right)+t(t-u) \left(\log\frac{\mu^2}{-t}+\frac{7}{6} \right)+ u(u-t)\left(\log\frac{\mu^2}{-u}+\frac{7}{6} \right)\right]\nn  \\
 & & + \cO(p^6)\,,
\eea}

\noi
where $l^{\rm r}_{1,2}$ are renormalized coupling constants of the $\cO(p^4)$ effective chiral Lagrangian at the scale $\mu$. The terms in the first line of Eq.~\rf{eq:GLAP} are leading in the QCD Large--$\Nc$ limit but so far, in this section, we are not restricting ourselves to this limit. The terms in the second line are induced by the chiral loops generated by the lowest order Lagrangian  renormalized at the scale $\mu$. The overall contribution of $\cO(p^4)$ is $\mu$--scale independent and well defined in the chiral limit. The relation between the $l_{i}^{\text r}$ constants and the more conventional $L_{i}^{\text r}$ constants of the  chiral $SU(3)$ Lagrangian~\cite{GL85} is as follows:

{\setl
\bea\lbl{eq:Llrels}
l_{1}^{\text r}(\mu) & = & 4 L_{1}^{\text r}(\mu) +2 L_{3}  -\frac{1}{96\pi^2}\frac{1}{8}\left(\log\frac{M_{K}^2}{\mu^2}+1 \right)\,, \\
l_{2}^{\text r}(\mu) & = & 4 L_{2}^{\text r}(\mu)  -\frac{1}{96\pi^2}\frac{1}{4}\left(\log\frac{M_{K}^2}{\mu^2}+1 \right)\,,
\eea}

\noi
where here, kaon particles have been treated as massive and  integrated out, hence the dependence on their mass $M_K$.

The linear combinations of the isospin amplitudes $T^I (s,0)$ which diagonalize the crossing matrix in the second line of Eq.~\rf{eq:roy} are:

{\setl
\bea
F_{1}(s,0) & = & -\frac{1}{6}\ T^0 (s,0) -\frac{1}{4}\ T^1 (s,0) +\frac{5}{12}\ T^2 (s,0)\,,\nn\\
F_{2}(s,0) & = & +\frac{1}{6}\ T^0 (s,0) +\frac{1}{4}\ T^1 (s,0) +\frac{7}{12}\ T^2 (s,0)\,,\nn\\
F_{3}(s,0) & = & -\frac{1}{6}\ T^0 (s,0) +\frac{3}{4}\ T^1 (s,0) +\frac{5}{12}\ T^2 (s,0)\,,
\eea}

\noi
and the physical elastic forward scattering amplitudes we are concerned with are then given by:

{\setl
\bea
\cA_{\pi^{\pm} \pi^0 \ra \pi^{\pm} \pi^0}(s,0) & = & \frac{1}{2}\left[F_{2}(s,0)+F_{3}(s,0)\right] =\frac{1}{2}\left[T^1 (s,0) +T^2 (s,0)\right]\,,\nn\\
\cA_{\pi^0 \pi^0 \ra \pi^0 \pi^0}(s,0) & = & \frac{1}{2}\left[ 3F_2 (s,0) -F_3 (s,0)\right] = \frac{1}{3}\left[T^0 (s,0) + 2\ T^2 (s,0)\right] \,,\nn\\
\cA_{\pi^{+} \pi^{-} \ra \pi^{+} \pi^{-}}(s,0) & = & -F_1 (s,0) +F_2 (s,0) = \frac{1}{3}T^0 (s,0) + \frac{1}{2}T^1 (s,0)+\frac{1}{6}T^2 (s,0)\,,\nn\\
\cA_{\pi^{\pm} \pi^{\pm} \ra \pi^{\pm} \pi^{\pm}}(s,0) & = & F_{1}(s,0)+F_{2}(s,0)= T^2 (s,0)\,.
\eea}

\noi
The Roy equations for the $F_{i} (s,0)$ amplitudes are then:

{\setl
\bea\lbl{eq:roydiag}
\Ree
\lefteqn{
\left(\begin{array}{c}
{F}_{1} (s,0)\\
{F}_{2} (s,0)\\
{F}_{3} (s,0)
\end{array}\right)=\frac{s}{f_{\pi}^2} \left(\begin{array}{c} -1 \\ ~~0 \\ ~~0
\end{array}~~\right)} \nn \\
& & +s^2 \int_0^{\infty}ds' \frac{1}{s'^2}\left[\frac{1}{s'-s}\left(\begin{array}{ccc} 1 & 0 & 0   \\
0 & 1  & 0  \\
0 & 0 & 1 \end{array} \right) +\frac{1}{s'+s}\left(\begin{array}{ccc} -1 & 0 & 0 \\
 0 & 1 & 0   \\
 0 & 0 & 1 \end{array} \right)\right]\frac{1}{\pi}\Imm \left(\begin{array}{c}
{F}_{1} (s',0)\\
{F}_{2} (s',0)\\
{F}_{3} (s',0)\end{array}\right)\,.
\eea}

\noi
From these equations there follows that the amplitudes $F_2$ and $F_3$  obey the same dispersion relation
\be\lbl{eq:F23}
\Ree F_{2,3}(s,0) = s^2 \int_0^\infty \frac{ds'^2}{s'^2}\frac{1}{s'^2 -s^2}\frac{1}{\pi}\Imm F_{2,3}(s',0)\,,
\ee
and  are even under $s\leftrightarrow -s$, while the amplitude $F_{1}(s,t)$ obeys the dispersion relation
\be\lbl{eq:f1}
\Ree F_1 (s,0)= -\frac{s}{f_{\pi}^2}+2 s^3 \int_0 ^\infty \frac{ds'}{s'^2}\frac{1}{s'^2 -s^2}\frac{1}{\pi}\Imm F_1 (s',0)\,,
\ee
and is odd under $s\leftrightarrow -s$. Indeed, one can check that there is no contribution of $\cO(s^2)$ to the $F_1 (s,0)$ amplitude in $\chi$PT, while the contributions of that order from $\chi$PT to the $F_{2}(s,0)$ and $F_{3}(s,0)$ amplitudes are:

{\setl
\bea\lbl{eq:chip4}
\Ree F_2 (s,0) & \underset{s \rightarrow 0}{=}\ & \frac{s^2}{f_{\pi}^4}\left[ 2l_1^{\rm r} +3l_2^{\rm r} +\frac{1}{12\pi^2}\left(\log\frac{\mu^2}{s}+\frac{25}{24}\right)\right] + \cO(s^4 )\,,\\
\Ree F_3 (s,0) & \underset{s \rightarrow 0}{=}\ & \frac{s^2}{f_{\pi}^4}\left[ -2l_1^{\rm r} +l_2^{\rm r} +\frac{1}{96\pi^2}\right] + \cO(s^4 )\,.
\eea}

\vspace*{0.5cm}

\section{\normalsize Mellin--Barnes Representation for the $F_i (s,0)$ Amplitudes.}
\setcounter{equation}{0}
\def\theequation{\arabic{section}.\arabic{equation}}

\noi
The optical theorem relates the amplitudes $\Imm F_{i}(s,0)$ to the total $\pi\pi$ cross sections as follows (massless pions):

{\setl
\bea
\Imm F_{1} (s,0)  & = & \frac{1}{2}\left[s\ \sigma_{\pi^{+}\pi^{+}}^\text{tot} -s\ \sigma_{\pi^{+}\pi^{-}}^\text{tot}\right]\,,\nn \\
\Imm F_{2} (s,0) & = & \frac{1}{2}\left[s\ \sigma_{\pi^{+}\pi^{+}}^\text{tot} +s\ \sigma_{\pi^{+}\pi^{-}}^\text{tot}\right]= \frac{1}{2}\left[s\ \sigma_{\pi^{+}\pi^{0}}^\text{tot} +s\ \sigma_{\pi^{0}\pi^{0
}}^\text{tot}\right]\,,\nn \\
\Imm F_{3} (s,0)& = & \frac{1}{2}\left[3s\ \sigma_{\pi^{+}\pi^{0}}^\text{tot} -s\ \sigma_{\pi^{0}\pi^{0}}^\text{tot}\right]\,.
\eea}

\noi
Let us then consider the Mellin transforms of the $\frac{1}{\pi}\Imm F_i (s,0)$ amplitudes
\be\lbl{eq:mellin}
\Sigma_{i}(\xi)=\int_0^\infty d\left(\frac{s}{M^2}\right) \left(\frac{s}{M^2}\right)^{\xi-1}\frac{1}{\pi}\Imm F_{i}\left(s,0\right)\,,
\ee 
and the corresponding  inverse Mellin transforms,
\be
\lbl{eq:ass}
\frac{1}{\pi}\Imm F_i (s,0)  =  \frac{1}{2\pi i}\int_{c-i\infty}^{c+i\infty} d\xi \left( \frac{s}{M^2}\right)^{-\xi}\Sigma_i (\xi)\,,
\ee
where, for convenience, we have introduced an arbitrary mass scale $M$ (e.g. the $\rho$ mass) so as to normalize the dimensions of the $s$ variable.
We then observe the following facts:

\begin{itemize}
	\item 
Acording to Eq.~\rf{eq:FM}, a FM--like asymptotic behaviour for the physical $\sigma_{\pi\pi}^{\text tot}(s)$ cross sections, implies

{\setl
\bea\lbl{eq:sigmaA}
\sigma^{\mbox{\footnotesize tot}}_{\pi^{+}\pi^{+}}(s)&  \underset{{s\ \ra \infty}}{\thicksim}\ & A_{\pi^{+}\pi^{+}}\ \frac{\pi }{M^2}\log^2 \frac{s}{M^2}\,,\nn\\
\sigma^{\mbox{\footnotesize tot}}_{\pi^{+}\pi^{-}}(s)&  \underset{{s\ \ra \infty}}{\thicksim}\ & A_{\pi^{+}\pi^{-}}\ \frac{\pi }{M^2}\log^2 \frac{s}{M^2}\,,\nn\\
\sigma^{\mbox{\footnotesize tot}}_{\pi^{+}\pi^{0}}(s)&  \underset{{s\ \ra \infty}}{\thicksim}\ & A_{\pi^{+}\pi^{0}}\ \frac{\pi }{M^2}\log^2 \frac{s}{M^2}\,,\nn\\
\lbl{eq:assigmaF4}\sigma^{\mbox{\footnotesize tot}}_{\pi^{0}\pi^{0}}(s)&  \underset{{s\ \ra \infty}}{\thicksim}\ & A_{\pi^{0}\pi^{0}}\ \frac{\pi }{M^2}\log^2 \frac{s}{M^2}\,,
\eea}

\noi
where the $A_{\pi\pi}$ are some appropriate constants. According to the normalization implied by Eq.~\rf{eq:FM} they should all be fixed to
\be
\lbl{eq:andremartin}
A_{\pi\pi}\big\vert_\text{FM}=\frac{M^2}{m_{\pi}^2}\,,
\ee
but here we consider  the  $A_{\pi\pi}$ constants as {\it a priori} unknown.

The {\it inverse mapping theorem}~\cite{FGD95} requires then that if the  asymptotic behaviours in Eqs.~\rf{eq:assigmaF4} are satisfied, the Mellin transforms of the $\frac{1}{\pi}\Imm F_{i}(s,0)$ amplitudes must have a triple pole at $\xi\ra -1$:
\be\lbl{eq:mellinsigma}
\Sigma_i (\xi) \underset{{\xi \ra -1}}{\thicksim}\ \frac{{ -} 2 a_i}{(\xi+1)^3}\,,
\ee
where

{\setl
\bea
a_1 & = & \frac{1}{2}\left[ A_{\pi^{+}\pi^{+}}- A_{\pi^{+}\pi^{-}}\right]\,,\nn \\
a_2 & = & \frac{1}{2}\left[ A_{\pi^{+}\pi^{+}}+ A_{\pi^{+}\pi^{-}}\right]
 = \frac{1}{2}\left[ A_{\pi^{+}\pi^{0}}+A_{\pi^{0}\pi^{0}}\right]\,,\nn \\
a_3 & = & \frac{1}{2}\left[3 A_{\pi^{+}\pi^{0}}- A_{\pi^{0}\pi^{0}}\right]\lbl{eq:iphys}\,,
\eea}

\noi
The leading singularity of $\Sigma_i (\xi)$ in the Mellin plane at the {\it right} of the  {\it fundamental strip} (which fixes the integration boundary $c$ in the inverse Mellin  transform in Eq.~\rf{eq:ass})  must then be at $\xi =-1$ and it must be a triple pole. If all the $A_{\pi\pi}$ constants are equal, then  $a_1 =0$ and the corresponding pole at $\xi =-1$ becomes, at most, a double pole. We assume however, for the sake of generality, that all the  $a_i \neq 0$.

\item
Let us next consider the Mellin--Barnes representation of the dispersion relations for the amplitudes $F_{i} (s,0)$ in Eq.~\rf{eq:roydiag}. Using the relations

{\setl
\bea\lbl{eq:david}
\frac{1}{1+A} & = & 
\frac{1}{2\pi i}\int\limits_{c_\xi-i\infty}^{c_\xi+i\infty}d\xi\ 
A^{-\xi}\  \Gamma(\xi)\Gamma(1-\xi)\,, \\
\frac{1}{1-A} & = & 
\frac{1}{2\pi i}\int\limits_{c_\xi-i\infty}^{c_\xi+i\infty}d\xi\ 
A^{-\xi}\ \Gamma(\xi)\Gamma(1-\xi)\  \frac{\pi}{\Gamma\left(\frac{1}{2}+\xi \right)\Gamma\left(\frac{1}{2}-\xi \right)}\,,
\eea}

\noi
and respecting the $s\leftrightarrow -s$ symmetry properties of the $\Ree {F}_{i} (s,0)$ amplitudes, one finds

\begin{align}
\lbl{eq:roybarnes}
&\Ree\left(\begin{array}{c}
{F}_{1} (s,0)\\
{F}_{2} (s,0)\\
{F}_{3} (s,0)
\end{array}\right)  = \frac{s}{f_{\pi}^2} \left(\begin{array}{c} -1 \\ ~~0 \\ ~~0
\end{array}~~\right) \nonumber \\
 &+ \frac{1}{2\pi i}\int\limits_{c_\xi-i\infty}^{c_\xi+i\infty}d\xi\, \Gamma(\xi)\Gamma(1-\xi) \times 
\begin{pmatrix}
\Sigma_1(\xi-2) \frac{s}{M^2}\left(\frac{|s|}{M^2}\right)^{1-\xi}\left[1-\frac{\pi}{\Gamma\left(\frac{1}{2}+\xi\right)\Gamma\left(\frac{1}{2}-\xi\right)}\right] \\
\Sigma_2(\xi-2)\left(\frac{|s|}{M^2}\right)^{2-\xi}\left[1+\frac{\pi}{\Gamma\left(\frac{1}{2}+\xi\right)\Gamma\left(\frac{1}{2}-\xi\right)}\right] \\
\Sigma_3(\xi-2)\left(\frac{|s|}{M^2}\right)^{2-\xi}\left[1+\frac{\pi}{\Gamma\left(\frac{1}{2}+\xi\right)\Gamma\left(\frac{1}{2}-\xi\right)}\right]
\end{pmatrix}
\end{align}

\noi
where we have used the fact that
\be
\int_0^\infty d\left(\frac{s'}{M^2} \right)\left(\frac{s'}{M^2} \right)^{\xi-3}
\frac{1}{\pi}\Imm F_{i}(s',0)  =  \Sigma_{i}(\xi-2)\,,
\ee
with $\Sigma_{i}(\xi)$ the same Mellin transform as the one defined in Eq.~\rf{eq:mellin}. Notice that the {\it fundamental strip} in Eq.~\rf{eq:roybarnes} is now defined by $c_{\xi}=\Ree (\xi) \in\ \rbrack 0,1\lbrack$. Again, the low energy behaviour of the  $F_{i}(s,0)$ amplitudes is  governed by the  singularities at  the {\it left} of this {\it fundamental strip}, while their high energy behaviour is governed by the singularities at the {\it right} of the same {\it fundamental strip}.

\item
In particular, the leading low energy behaviours of the $F_{i} (s,0)$ amplitudes are governed by the values of the $\Sigma_{i}(\xi-2)$ at $\xi\ra 0$ and leads to the results:

{\setl
\bea\lbl{eq:mellinchi}
\Ree F_{1} (s,0) & \underset{{s\ \ra  0}}{=}\  &  -\frac{s}{f_{\pi}^2}+ \cO(s^3)\,,\\
\Ree F_{2} (s,0) & \underset{{s\ \ra  0}}{=}\ &  \frac{s^2}{M^4}\ \lim_{\xi \rightarrow 0} \left\{\frac{d}{d\xi}\left[2 \xi \Sigma_{2}(\xi-2)\right] \log \frac{M^2}{s} + 2 \xi \Sigma_{2}(\xi-2) \right\}\! +\!\cO(s^4)\,,\\
\Ree F_{3} (s,0) & \underset{{s\ \ra  0}}{=}\ &  \frac{s^2}{M^4}\ 2\Sigma_{3}(-2) +\cO(s^4)\,.
\eea}

\noi
Comparison with the $\chi$PT expansion in Eqs.~\rf{eq:chip4} allows us then to fix the values of {$\Sigma_{2,3}$ at $\xi=-2$} to

{\setl
\bea\lbl{eq:mellinfc}
\Sigma_2(\xi) & \underset{{\xi\ \ra  -2}}{\thicksim}\ & \frac{M^4}{f_{\pi}^4} \left[
l_1^{\rm r} +\frac{3}{2}l_2^{\rm r}  +\frac{25}{576\pi^2}\right] + \frac{1}{24\pi^2}\frac{1}{\xi+2}\,, \\
\Sigma_3(\xi) & \underset{{\xi\ \ra  -2}}{\thicksim}\ & \frac{M^4}{f_{\pi}^4}\left[ -l_1^{\rm r} +\frac{1}{2}l_2^{\rm r} +\frac{1}{192\pi^2}\right] \,. 
\eea}

\noi

\item
On the other hand, the leading high energy behaviours of the $F_{i} (s,0)$ amplitudes are governed by the  $\Sigma_{i}(\xi-2)$ at $\xi\ra 1$ which, if the FM bound is saturated for all the $\sigma^{\mbox{\footnotesize tot}}_{\pi\pi}$ cross sections, have triple poles at the values:
\be
\Sigma_{i} (\xi -2)  \underset{{\xi\ \ra  1}}{\thicksim}\ \frac{ { -}2 a_{i}}{(\xi-2+1)^3}\,.
\ee
For the amplitudes $\Ree F_2 (s,0)$ and $\Ree F_3 (s,0)$ the effect of this triple pole is softened by the fact that
\be
\frac{\pi}{\Gamma(\frac{1}{2}+\xi)\Gamma(\frac{1}{2}-\xi)}\underset{{\xi\ \ra  1}}{\thicksim}\ -1+\frac{\pi^2}{2}(\xi-1)^2 +\cO\left(\xi -1\right)^4 \,,
\ee
and there is a cancellation between the two  terms in the brackets in the second line at the r.h.s. of Eq.~\rf{eq:roybarnes}.
Therefore, the leading asymptotic behaviour of $ F_{2,3} (s,0)$ is then of the type
\be
\Ree F_{2,3} (s,0)  \underset{{s\ \ra  \infty}}{\thicksim}\ \cO[a_{2,3} \ \vert s\vert\log  \vert s\vert]\,.
\ee

By contrast, if $a_1 \neq 0$, there is no such a cancellation for the $F_1 (s,0)$ amplitude  and its leading high energy behaviour  will then be of the type:
\be
\Ree F_{1}(s,0)  \underset{{s\ \ra  \infty}}{\thicksim}\ -\frac{s}{f_{\pi}^2} + \cO[a_1 \ s\log^3 \vert s\vert]\,,
\ee

\end{itemize}

From the previous considerations we conclude that the Mellin--Barnes representation of the elastic  $\pi\pi$ forward scattering amplitudes $F_i (s,0)$ show explicitly how their asymptotic behaviours for $s\ra\infty$ (relevant to the FM bound), and for $s\ra 0$ (relevant to the $\chi$PT expansion), are governed by  the Mellin transforms $\Sigma_{i}(\xi)$ defined in Eq.~\rf{eq:mellin} . We find from $\chi$PT that the chiral limits ($m_{\pi}\ra 0$) of these Mellin functions exist and are perfectly well defined in QCD at the {\it left} of the corresponding {\it fundamental strips}. We  have also shown how the high energy behaviours of the $F_i (s,0)$ amplitudes, are governed by the same Mellin transforms and, therefore, the FM bound has direct implications on their behaviours. If, as implied by the normalization of the FM bound in Eq.~\rf{eq:FM}, and hence Eq.~\rf{eq:andremartin},  the Mellin functions $\Sigma_{i}(\xi)$ at the {\it right} of their {\it fundamental strips} are singular in the chiral limit, it means that they must have a discontinous behaviour with respect to the pion mass in the sense that: {\it they exist in the chiral limit  at the left of their fundamental strips yet they blow up to infinity, in the same limit, at the right of their fundamental strips.} This we find a rather peculiar behaviour which, although mathematically possible, questions the presence of a pion mass factor in the denominator of the normalization of the FM bound in QCD.

\vspace*{0.5cm}

\section{\normalsize The Froissart--Martin Bound in the QCD Large--$\Nc$ Limit.}
\setcounter{equation}{0}
\def\theequation{\arabic{section}.\arabic{equation}}

\noi
In this section   we shall directly work   with the $\pi\pi$ scattering  amplitudes $\Imm T^{I}(s,0)$ with well defined isospin ($I=0,1,2$). They are related to the $\Imm F_{i}(s,0)$ amplitudes which we have considered in the previous section as follows:

{\setl
\bea\lbl{eq:isosF}
\Imm T^0 (s,0) & = & \frac{1}{2}\left[-4\Imm F_1 (s,0)+5\Imm F_2 (s,0)-3 \Imm F_3 (s,0)\right]\,,\nn\\
\Imm T^1 (s,0) & = & -\Imm F_1 (s,0)+\Imm F_3 (s,0)\,,\nn\\
\Imm T^2 (s,0) & = & \Imm F_1 (s,0)+\Imm F_2 (s,0)\,.
\eea}

\noi
In the Large--$\Nc$ limit of QCD, the $\Imm T^{I}(s,0)$ amplitudes   are composed of an infinite set of narrow states:
\be\lbl{eq:largeansatz}
\frac{1}{\pi}\Imm T^{I}(s,0)=\sum_{n=0}^{\infty}\vert F_{I,n}\vert^2 \delta\left(s-M_{I,n}^2\right)\,,\quad I=0,1,2\,.
\ee
The question is then the following:
{\it is it possible to find constraints on the couplings $F_{I,n}$ and the masses $M_{I,n}$ of a possible Large--$\Nc$ ansatz
so as to reproduce the FM asymptotic behaviour for the $\sigma^{\mbox{\footnotesize tot}}_{\pi\pi}$ cross sections?}

In order to answer this question we shall proceed as follows. 
The Mellin transforms of $\frac{1}{\pi}\Imm T^{I}(s,0)$ in the Large--$\Nc$ limit are given by Dirichlet--like series~\footnote{For a recent discussion of QCD Large--$\Nc$ properties in connexion with the asymptotic behaviours of two--point functions see ref.~\cite{EdeR12}.}:
\be\lbl{eq:mellinNc}
\Sigma^{I}(\xi) = \sum_{n=0}^{\infty}\frac{\vert F_{I,n}\vert^2}{M^2}\left( \frac{M^2}{M_{I,n}^2}\right)^{-\xi+1}\,,
\ee 
and the corresponding  inverse Mellin transforms are:
\be\lbl{eq:assNc}
\frac{1}{\pi}\Imm T^{I}(s,0) =
 \frac{1}{2\pi i}\int_{c-i\infty}^{c+i\infty} d\xi \left( \frac{s}{M^2}\right)^{-\xi} \sum_{n=0}^{\infty}\frac{\vert F_{I,n}\vert^2}{M^2}\left( \frac{M^2}{M_{I,n}^2}\right)^{-\xi+1} \,.
\ee
As discussed in the previous section, a FM--like asymptotic behaviour for the $\sigma^{\mbox{\footnotesize tot}}_{\pi\pi}$ cross sections fixes the leading singularity of the Mellin transforms of the $\frac{1}{\pi}\Imm F_{i}(s,0)$ amplitudes as given in Eqs.~\rf{eq:mellinsigma} and \rf{eq:iphys} and therefore, from Eqs.~\rf{eq:isosF}, there follows that 
\be\lbl{eq:poleI}
\Sigma^{I} (\xi)  \underset{{\xi\ \ra  -1}}{\thicksim}\ \frac{{ -} 2 A^{I}}{(\xi+1)^3}\,,
\ee
where

{\setl
\bea
A^0 & = & \frac{1}{2}\left[-4a_1 +5a_2 -3 a_3   \right]\,,\nn\\
A^1 & = & -a_1 +a_3 \,,\nn\\
A^2 & =& a_1 +a_2\,.\lbl{eq:Aapi}
\eea}

\noi

In order to  construct a  simple Large--$\Nc$  {\it ansatz} with the required properties  we shall assume a Regge growth for the masses of the narrow states with $I=1$:
\be\lbl{eq:regge}
M_{I=1,n}^2 = M_{\rho}^2 +n\Lambda^2\,, 
\ee
and the absence of exotic trajectories i.e., no poles with $I=2$. We are then assuming that  the $I=1$ channel fully dominates the physical 
$\cA[\pi^{\pm} \pi^0 \ra \pi^{\pm} \pi^0]$ amplitude  and  focus our attention on this amplitude in the limit where 
\be\lbl{eq:pipivd}
\cA_{\pi^{\pm}\pi^0 \ra  \pi^{\pm}\pi^0}(s,0)=\frac{1}{2}\left[F_{2}(s,0)+F_{3}(s,0) \right]\simeq \frac{1}{2}T^{I=1} (s,0)\,.
\ee
Saturation of the FM  bound for the corresponding total cross section $\sigma_{\pi^{\pm}\pi^0}^{{\mbox{\footnotesize tot}}}$ requires  the couplings $\frac{\vert F_{I=1,n}\vert^2}{M^2}$ in Eq.~\rf{eq:mellinNc} to grow like $n \log^2 n$  as $n\ra\infty$.
The simplest form of a Large-$\Nc$ $\Imm T^{I=1} (s,0)$ amplitude satisfying these requirements is then 
\be\lbl{eq:constc}
\frac{1}{\pi}\Imm T^{I=1}(s,0)={\rm C} \sum_{n=0}^\infty 
\left(M_{\rho}^2 +n\Lambda^2 \right) \log^2 \left(\frac{M_{\rho}^2}{\Lambda^2} +n \right)\delta(s-M_{\rho}^2 -n\Lambda^2 )\,,
\ee
where ${\rm C}$ denotes  a dimensionless constant. Here we have fixed the arbitrary scale $M$ to $M =M_{\rho}$,   and $\Lambda$ is the mass scale which as shown in Eq.~\rf{eq:regge} fixes the equally spaced Regge--like spectra.
Quite remarkably, the Mellin transform of this Large--$\Nc$ {\it  ansatz} has a close analytic form:
\be\lbl{eq:truemellin}
\Sigma^{I=1}(\xi)=    {\rm C} \left(\frac{M_{\rho}^2}{\Lambda^2}\right)^{-\xi} { \frac{d^2}{d\xi^2}}\ \zeta\left( -\xi,\frac{M_{\rho}^2}{\Lambda^2}\right)\,,
\ee
where $\zeta\left( -\xi,\frac{M_{\rho}^2}{\Lambda^2}\right)$ is the Hurwitz function, a generalization of the Riemann zeta function, defined by the series:
\be
\zeta(\xi,v)=\sum_{n=0}^{\infty}\frac{1}{(n+v)^{\xi}}\,,\quad { \Ree \, \xi>1 ,\quad \text{with $\quad 0 < v \leq 1$ }},
\ee
and its analytic continuation. For $v=1$ it reduces to the Riemann zeta function. The asymptotic behaviour of $\frac{1}{\pi}\Imm T^{I=1}(s,0)$ for $s\ra\infty$ and hence of $\sigma_{\pi^+ \pi^0}^{\text{tot}}$, is governed by the residue of the triple pole of  $\Sigma^{I=1}(\xi)$ at $\xi =-1$ (see Eq.~\rf{eq:poleI}). This relates the overall constant ${\rm C} $ in Eq.~\rf{eq:constc} to the coefficient $A_{\pi^+ \pi^0}$ in Eq~\rf{eq:sigmaA} and hence:
\be
\sigma^{\mbox{\footnotesize tot}}_{\pi^{+}\pi^{0}}(s) \underset{{s\ \ra \infty}}{\thicksim}\  \frac{{\rm C} }{2}\left(\frac{M_{\rho}^2}{\Lambda^2} \right)\ \frac{\pi }{M_{\rho}^2}\log^2 \frac{s}{M_{\rho}^2}\,.
\ee

Let us next discuss the low--energy constraint that we can impose to the simple large--$\Nc$ {\it  ansatz} in Eq.~\rf{eq:constc} so as to fix the value of the overall constant ${\rm C}$. The isospin $I=1$ dominance assumption of the $\chi$PT expressions in Eqs.~\rf{eq:mellinfc}, when restricted to the Large--$\Nc$ limit, fixes them to the values

{\setl
\bea\lbl{eq:mellinfclarge}
\Sigma_2(\xi) & \underset{{\xi\ \ra  -2}}{\thicksim}\ & \frac{M^4}{f_{\pi}^4} \left[
l_1 +\frac{3}{2}l_2\right]\sim \frac{1}{4}\frac{M_{\rho}^2}{f_{\pi}^2} \,, \\
\Sigma_3(\xi) & \underset{{\xi\ \ra  -2}}{\thicksim}\ & \frac{M^4}{f_{\pi}^4}\left[ -l_1 +\frac{1}{2}l_2 \right]\sim \frac{3}{4}\frac{M_{\rho}^2}{f_{\pi}^2} \,, 
\eea}

\noi
where in the r.h.s. we have used the $\rho$--dominance approximation for the $l_i$ constants~\cite{EGPdeR89,EGLPdeR89} and, as before, we have fixed $M=M_{\rho}$.
Matching the Large--$\Nc$ {\it  ansatz} in Eq.~\rf{eq:truemellin} to these considerations, using Eq.~\rf{eq:pipivd},  fixes the ${\rm C}$ constant to
\be
{\rm C}\sim \frac{1}{ \zeta''\left(2,\frac{M_{\rho}^2}{\Lambda^2}\right)}\ \frac{\Lambda^4}{f_{\pi}^2 M_{\rho}^2}\,,
\ee
and, therefore, the leading asymptotic growth of the total $\pi^+ \pi^0$ cross section to:
\be\lbl{eq:ncfin}
\sigma^{\mbox{\footnotesize tot}}_{\pi^{\pm}\pi^{0}}(s)
 \underset{{s\ \ra \infty}}{\thicksim}\ \frac{\frac{\Lambda^2}{M_{\rho}^2}}{2\  \zeta''\left(2,\frac{M_{\rho}^2}{\Lambda^2}\right)}\frac{\pi}{f_{\pi}^2 }
 \log^2 \frac{s}{M_{\rho}^2} \,.
 \ee
Like the usual FM bound, it grows as $\log^2 s$, but it is finite in the chiral limit and of $\cO(1/\Nc)$ in the Large--$\Nc$ counting. Numerically, for $\Lambda =M_{\rho}$,  the r.h.s. of Eq~\rf{eq:ncfin} becomes
\be
\sigma^{\mbox{\footnotesize tot}}_{\pi^{\pm}\pi^{0}}(s)
 \underset{{s\ \ra \infty}}{\thicksim}\ 0.25\ \frac{\pi}{f_{\pi}^2}
  \log^2 \frac{s}{M_{\rho}^2} \,,
 \ee
 but we should stress that this is only a Large--$\Nc$ model estimate with many simplifications.

\vspace*{0.5cm}

\section{\normalsize Conclusions}
\setcounter{equation}{0}
\def\theequation{\arabic{section}.\arabic{equation}}

\noi
We have shown that it is possible to construct a Large--$\Nc$ QCD {\it ansatz} compatible with the Froissart--Martin bound. The bound, however, is finite in the chiral limit and it is of $\cO(1/\Nc)$ in the Large--$\Nc$ counting rules. In fact, it seems very likely that these two features should be generic to full QCD because of the
fact that QCD has spontaneous chiral symmetry breaking. This implies the existence of a mass gap between the Nambu--Goldstone states (the pions) and the other hadronic states. It is this property which, very likely,  forces the presence of characteristic scales like $f_{\pi}$ in the normalization of the FM bound and which at the same time provides the  correct $\cO (1/\Nc)$ Large--$\Nc$ counting. 

The usual derivation of the FM bound does not take into account  the fact that the underlying dynamics of the strong interactions has the property of spontaneous chiral symmetry breaking. In fact, it implicitly assumes a realization  of the hadronic spectrum a la Wigner--Weyl without Nambu--Goldstone particles, in which case, the normalization in Eq.~\rf{eq:FM} is not surprising.

Finally, we wish to emphasize that the discussion above does not answer the third question in the introduction. We have only shown that, in the Large--$\Nc$ limit of QCD,  it is possible to construct models which  {\it a priori}  show no obstruction for the asymptotic behaviour of the total $\pi\pi$ cross sections to saturate a $\log^2 s$ like behaviour.

\newpage
\begin{center}
{\normalsize\bf Acknowledgments.}
\end{center}
\vspace*{0.25cm}

We are very grateful to Gerhard Ecker and Marc Knecht for useful discussions and their help with the properties of elastic $\pi\pi$ scattering at low energies.  An encouraging conversation of one of us (EdeR) with G\"unter Dosch at the early stages of our interest on this subject is also acknowledged.
 
\vspace*{1.2cm}


\end{document}